\documentclass[journal]{IEEEtran}

\usepackage[utf8]{inputenc}
\usepackage[T1]{fontenc}
\usepackage{amsmath,amssymb,amsthm}
\usepackage{cite}
\usepackage{bm}
\usepackage{mathtools}
\usepackage{url}
\usepackage[hidelinks]{hyperref}

\newtheorem{theorem}{Theorem}
\newtheorem{lemma}{Lemma}
\newtheorem{proposition}{Proposition}
\newtheorem{corollary}{Corollary}
\newtheorem{remark}{Remark}

\newcommand{\Acal}{\mathcal{A}}
\newcommand{\Wcal}{\mathcal{W}}

\newcommand{\E}{\mathbb{E}}
\newcommand{\Pp}{\mathbb{P}}
\newcommand{\Var}{\operatorname{Var}}
\newcommand{\Cov}{\operatorname{Cov}}
\newcommand{\Normal}{\mathcal{N}}
\newcommand{\ind}{\mathbf{1}}
\newcommand{\dd}{\mathrm{d}}
\newcommand{\norm}[1]{\left\lVert #1\right\rVert}

\title{\LARGE Predictive Self-Entropy in Dirichlet Universal Coding:\\ A Factor-Two Central Limit Theorem and a Markov Extension}

\author{Thomas Sch\"urmann\\Düsseldorf, Germany%
\thanks{Thomas Sch\"urmann is an independent researcher. Email: \url{t.schurmann@icloud.com}. ORCID: \url{0000-0003-1928-5213}.}}

\begin{document}
\maketitle

\begin{abstract}
	Sequential universal coders produce both a realized log-loss and, before each symbol is observed, a predictive distribution with its own Shannon entropy. We compare the realized arithmetic code length with the cumulative predictive self-entropy, the prequential sum of the self-expected logarithmic scores of the issued forecasts. This forecast-only quantity is neither a plug-in entropy estimate nor a posterior entropy functional. For finite-alphabet memoryless sources and Dirichlet (add-$\alpha$) sequential predictors, we prove a joint central limit theorem for the two normalized quantities. They have the same first-order entropy limit but different second-order fluctuations: for every nonuniform i.i.d.\ source, cumulative predictive self-entropy has exactly twice the leading asymptotic variance of realized code length, and the ordinary mean-square-error ratio converges to two. We also give its logarithmic bias coefficient. For fixed finite-order Markov sources with context-wise Dirichlet predictors, the factor-two result becomes a context/innovation covariance decomposition, recovering the i.i.d.\ factor two when only the innovation component contributes.
\end{abstract}


\IEEEpeerreviewmaketitle

\section{Introduction}
Sequential probability assignment is a common mathematical core of universal prediction, universal lossless compression, and statistical estimation.  In the prequential framework of Dawid \cite{Dawid1984,Dawid1999}, forecasts are evaluated in the order in which they are issued.  Under log-loss this evaluation has a direct coding interpretation: if a sequential predictor assigns a distribution \(Q_t\) to the next symbol before \(X_t\) is observed, then
\begin{equation}\label{eq:general-log-loss}
L_n=-\sum_{t=1}^n \log Q_t(X_t)
\end{equation}
is the ideal arithmetic code length of the observed string under that predictor, in nats.  In information theory, this viewpoint underlies the classical connections among universal prediction, universal coding, redundancy, and asymptotic likelihood theory; see, for example, Krichevsky and Trofimov \cite{KrichevskyTrofimov1981}, Rissanen \cite{Rissanen1984}, Clarke and Barron \cite{ClarkeBarron1990}, Merhav and Feder \cite{MerhavFeder1998}, and Xie and Barron \cite{XieBarron2000}.

The same sequential coder also computes, before seeing \(X_t\), the entire distribution \(Q_t\).  Hence it also computes the Shannon entropy \(H(Q_t)\).  We study the cumulative predictive self-entropy
\begin{equation}\label{eq:general-self-entropy}
S_n=\sum_{t=1}^n H(Q_t).
\end{equation}
This quantity is neither a second realized code length nor a final-time or posterior entropy functional: it depends on the next realized symbol only through the past data that determined \(Q_t\).  The adjective ``self'' emphasizes that the expectation is taken under the quoted predictive distribution itself, not under a Dirichlet posterior distribution or a posterior expectation of the source entropy.

The distinction between \eqref{eq:general-log-loss} and \eqref{eq:general-self-entropy} is easy to overlook.  The log-loss \(L_n\) is the ex-post code length paid on the realized data, whereas \(S_n\) is the ex-ante uncertainty reported by the same sequential predictor along the same path.  Since \(H(Q_t)\) averages over the quoted forecast rather than over the next realized symbol, one might expect this forecast-internal quantity to be at least as stable as realized log-loss when both are normalized as entropy estimators.  One purpose of this paper is to show that this intuition is false at the central-limit scale.

This forecast-internal interpretation also connects \eqref{eq:general-self-entropy} to the theory of proper scoring rules.  The entropy, or generalized entropy, associated with a scoring rule is the expected score obtained when the quoted distribution is the data-generating distribution; see Good \cite{Good1952}, Savage \cite{Savage1971}, Gneiting and Raftery \cite{GneitingRaftery2007}, and Gr\"unwald and Dawid \cite{GrunwaldDawid2004}.  For the logarithmic score this self-expected score is exactly Shannon entropy,
\begin{equation}\label{eq:self-score}
H(Q_t)=\E_{Y\sim Q_t}\big[-\log Q_t(Y)\big].
\end{equation}
Thus \(H(Q_t)\) is the generalized entropy of the forecast \(Q_t\), and \(S_n\) is its prequential accumulation along the prediction path.

Accordingly, \(S_n/n\) is not proposed here as a more efficient entropy estimator.  The question is instead whether realized prequential log-loss and a natural forecast-internal uncertainty functional produced by the same universal coder remain equivalent beyond first order.  For finite-alphabet memoryless sources and arbitrary strictly positive Dirichlet pseudo-counts, the answer is no.  The normalized quantities \(L_n/n\) and \(S_n/n\) have the same entropy limit, but their central-limit fluctuations differ: if the source is nonuniform and \(V(p)\) denotes its varentropy, then \(L_n/n\) has asymptotic variance \(V(p)/n\), whereas \(S_n/n\) has asymptotic variance \(2V(p)/n\).  Consequently, their ordinary mean-square-error ratio tends to two.

We also extend the comparison to fixed finite-order Markov sources with context-wise Dirichlet predictors.  In that setting the exact factor two is replaced by a structural covariance decomposition, separating fluctuations of the active context from the information innovations within that context.

\subsection{Relation to prior work}
The results are adjacent to, but distinct from, several well-developed literatures.  Classical entropy estimation for finite alphabets studies one-sample functionals such as \(H(\widehat p_n)\), bias-corrected plug-in estimators, or Bayesian posterior means of \(H(\theta)\).  The bias and asymptotic normality of the plug-in estimator go back at least to Miller \cite{Miller1955} and Basharin \cite{Basharin1959}; modern treatments and nonparametric refinements include Antos and Kontoyiannis \cite{AntosKontoyiannis2001} and Paninski \cite{Paninski2003}.  Bayesian and Dirichlet-based estimators of entropy as a functional of an unknown discrete distribution are developed, for example, by Wolpert and Wolf \cite{WolpertWolf1995} and in related work such as Scutari \cite{Scutari2018}.  These objects are final-time statistical functionals or posterior functionals, not the pathwise sum \(\sum_{t\le n}H(Q_t)\) generated by a sequential arithmetic predictor.

A second line of work uses compression algorithms to estimate entropy rates.  Earlier compression-based entropy-estimation methods for symbolic sequences, including finite-size extrapolation from code lengths, were studied by Sch\"urmann and Grassberger \cite{SchuermannGrassberger1996}.  The context-tree weighting method of Willems, Shtarkov, and Tjalkens \cite{WillemsShtarkovTjalkens1995} produces sequential mixture probabilities for finite-memory sources, and CTW-type code lengths have been used as practical entropy-rate estimators; see Gao, Kontoyiannis, and Bienenstock \cite{GaoKontoyiannisBienenstock2008}.  Bayesian Context Trees provide exact Bayesian inference for variable-memory Markov chains and yield posterior distributions of entropy rates and asymptotic normality statements for posterior entropy-rate quantities \cite{KontoyiannisEtAl2022BCT,PapageorgiouKontoyiannis2023ITW}.  The central object in these works is usually a realized universal code length, a posterior functional, or a model-based entropy-rate estimate.  The present paper instead studies the cumulative self-entropy of the forecasts issued along the prediction path and its joint fluctuations with realized prequential log-loss.

The contribution is focused on this prequential pair.  Conceptually, cumulative predictive self-entropy is identified as the logarithmic-score generalized entropy accumulated along a universal coding path.  Analytically, the paper proves a joint CLT with realized arithmetic code length, gives the logarithmic bias coefficient of \(S_n\), and extends the comparison to fixed finite-order Markov sources.

\section{Model and Definitions}\label{sec:model}
Let \(\Acal=\{1,\ldots,m\}\), \(m\ge 2\), and let \(X_1,X_2,\ldots\) be independent and identically distributed with distribution \(p=(p_1,\ldots,p_m)\), where \(p_i>0\) for all \(i\).  All logarithms are natural.  The Shannon entropy and varentropy are
\begin{equation}\label{eq:entropy}
H(p)=-\sum_{i=1}^m p_i\log p_i
\end{equation}
and
\begin{equation}\label{eq:varentropy}
V(p)=\Var_p[-\log p_X]=\sum_{i=1}^m p_i\bigl(-\log p_i-H(p)\bigr)^2,
\end{equation}
where \(X\sim p\).  Thus \(V(p)>0\) if and only if \(p\) is not uniform on its support, which here is the whole alphabet.

Fix a Dirichlet pseudo-count vector \(\alpha=(\alpha_1,\ldots,\alpha_m)\) with \(\alpha_i>0\) for all \(i\), and put \(\alpha_0=\sum_i\alpha_i\).  Let
\begin{equation}\label{eq:counts}
N_i(t)=\sum_{s=1}^t \ind\{X_s=i\}.
\end{equation}
For \(k\ge1\), write
\begin{equation}\label{eq:empirical-def}
\widehat p_k(i)=\frac{N_i(k)}{k},\qquad 1\le i\le m.
\end{equation}
The Dirichlet, or add-\(\alpha\), sequential predictor is
\begin{align}\label{eq:dirichlet-predictor}
q_t(i)&=\Pp_\alpha(X_t=i\mid X_1^{t-1}) \notag\\
&=\frac{N_i(t-1)+\alpha_i}{t-1+\alpha_0},\qquad 1\le i\le m.
\end{align}
Here \(\Pp_\alpha\) denotes only the Bayesian predictive assignment induced by the pseudo-counts \(\alpha\).  Unless explicitly stated otherwise, probabilities and expectations with subscript \(p\), such as \(\Pp_p\) and \(\E_p\), are taken under the true i.i.d.\ source.

The corresponding arithmetic code length and cumulative predictive self-entropy are
\begin{equation}\label{eq:L-alpha}
L_n^\alpha=-\sum_{t=1}^n \log q_t(X_t)
\end{equation}
and
\begin{equation}\label{eq:S-alpha}
S_n^\alpha=\sum_{t=1}^n H(q_t)=-\sum_{t=1}^n\sum_{i=1}^m q_t(i)\log q_t(i).
\end{equation}
When no confusion is possible, the superscript \(\alpha\) is suppressed.  We write \(Y_n=O_{L^r}(b_n)\) if \(\norm{Y_n}_{L^r}=(\E|Y_n|^r)^{1/r}=O(b_n)\).

The quantity \(S_n^\alpha\) is available online from the same state variables as the Dirichlet predictor.  If \(c_i=N_i(t-1)+\alpha_i\) and \(T=t-1+\alpha_0\), then
\begin{equation}\label{eq:online-entropy}
H(q_t)=\log T-\frac{1}{T}\sum_{i=1}^m c_i\log c_i.
\end{equation}
Thus the coder can add \(H(q_t)\) before observing \(X_t\).  If \(U_t=\sum_i c_i\log c_i\), then after observing \(X_t=x\) it updates, with \(c_x\) denoting the pre-update value,
\begin{equation}\label{eq:online-update}
U_{t+1}=U_t-c_x\log c_x+(c_x+1)\log(c_x+1),
\end{equation}
then sets \(c_x\leftarrow c_x+1\) and \(T\leftarrow T+1\).  Hence each update is \(O(1)\), and cumulative predictive self-entropy is a direct online by-product of the sequential arithmetic prediction path.

\section{Main Results}\label{sec:main}
Our main theorem gives the joint second-order behavior of \(L_n^\alpha/n\) and \(S_n^\alpha/n\).

\begin{theorem}[Joint CLT and factor-two MSE]\label{thm:main}
Assume that \(p_i>0\) for all \(i\) and that \(V(p)>0\); with full support, this is equivalent to nonuniformity of \(p\).  For every fixed Dirichlet pseudo-count vector \(\alpha=(\alpha_1,\ldots,\alpha_m)\) with \(\alpha_i>0\) for all \(i\),
\begin{equation}\label{eq:main-clt}
\sqrt n\left(\frac{L_n^\alpha}{n}-H(p),\frac{S_n^\alpha}{n}-H(p)\right)
\Rightarrow
\Normal\left(0,V(p)
\begin{pmatrix}
1&1\\[1mm]
1&2
\end{pmatrix}\right).
\end{equation}
Moreover,
\begin{align}
 n\E_p\left[\left(\frac{L_n^\alpha}{n}-H(p)\right)^2\right]&\longrightarrow V(p), \label{eq:mse-L}\\
 n\E_p\left[\left(\frac{S_n^\alpha}{n}-H(p)\right)^2\right]&\longrightarrow 2V(p). \label{eq:mse-S}
\end{align}
Consequently,
\begin{equation}\label{eq:mse-ratio}
\frac{\E_p\big[(S_n^\alpha/n-H(p))^2\big]}
{\E_p\big[(L_n^\alpha/n-H(p))^2\big]}
\longrightarrow 2.
\end{equation}
\end{theorem}

Theorem~\ref{thm:main} is invariant under logarithmic-order bias corrections.  Thus subtracting the usual parametric logarithmic redundancy from \(L_n^\alpha\), or subtracting an oracle or estimated logarithmic bias correction from \(S_n^\alpha\), changes lower-order bias terms but not the leading \(n^{-1}\) mean-square-error constants.

\begin{corollary}[Logarithmic corrections]\label{cor:corrections}
Let \(B_n\) and \(C_n\) be real random variables satisfying \(B_n=O_{L^2}(\log n)\) and \(C_n=O_{L^2}(\log n)\).  Under the assumptions of Theorem~\ref{thm:main},
\begin{equation}\label{eq:corrected-clt}
\sqrt n\left(\frac{L_n^\alpha-B_n}{n}-H(p),
\frac{S_n^\alpha-C_n}{n}-H(p)\right)
\end{equation}
has the same weak limit as in \eqref{eq:main-clt}.  The leading MSE constants remain \(V(p)\) and \(2V(p)\).
\end{corollary}

The logarithmic bias of \(S_n^\alpha\) differs qualitatively from the familiar redundancy coefficient of \(L_n^\alpha\).

\begin{proposition}[Logarithmic bias coefficients]\label{prop:bias}
Assume that \(p_i>0\) for all \(i\).  Then
\begin{align}
\E_p\big[L_n^\alpha-nH(p)\big]&=\frac{m-1}{2}\log n+O(1),\label{eq:bias-L}\\
\E_p\big[S_n^\alpha-nH(p)\big]&=b_S(p,\alpha)\log n+O(1),\label{eq:bias-S}
\end{align}
where
\begin{equation}\label{eq:bias-coeff}
b_S(p,\alpha)=-\sum_{i=1}^m(\alpha_i-\alpha_0p_i)\log p_i-\frac{m-1}{2}.
\end{equation}
In contrast to the logarithmic redundancy coefficient in \eqref{eq:bias-L}, the coefficient in \eqref{eq:bias-S} depends on both the source distribution \(p\) and the Dirichlet pseudo-count vector \(\alpha\).
\end{proposition}

\begin{remark}[Uniform sources]\label{rem:uniform}
If \(p_i=1/m\) for all \(i\), then \(-\log p_X\) is constant and \(V(p)=0\).  In this case the gradients of entropy in the simplex directions are constant and hence vanish after imposing \(\sum_i u_i=0\); equivalently, the linear variable \(\phi(X_t)\) in \eqref{eq:phi} is identically zero.  Lemma~\ref{lem:linear} then gives
\begin{align*}
L_n^\alpha-n\log m&=O_{L^2}(\log n),\\
S_n^\alpha-n\log m&=O_{L^2}(\log n),
\end{align*}
so the \(\sqrt n\)-normalization in Theorem~\ref{thm:main} converges to zero.  Proposition~\ref{prop:bias} also shows that the logarithmic means are of opposite sign,
\begin{align*}
\E_p[L_n^\alpha-n\log m]&=\frac{m-1}{2}\log n+O(1),\\
\E_p[S_n^\alpha-n\log m]&=-\frac{m-1}{2}\log n+O(1),
\end{align*}
independently of \(\alpha\) to logarithmic order.  A non-degenerate uniform-source theory would therefore have to center logarithmic terms and analyze quadratic Taylor components rather than the first-order information variables.  The factor-two comparison in Theorem~\ref{thm:main} is consequently a nonuniform-source phenomenon.
\end{remark}

\section{Linear Representations}\label{sec:linear}
The proof of Theorem~\ref{thm:main} rests on two asymptotic linear representations.  Define
\begin{equation}\label{eq:phi}
\phi(X_t)=-\log p_{X_t}-H(p),\qquad t\ge 1.
\end{equation}
Then \(\E_p\phi(X_t)=0\) and \(\E_p\phi(X_t)^2=V(p)\).  For \(1\le j\le n\), set
\begin{equation}\label{eq:weights}
a_{n,j}=\sum_{t=j+1}^n\frac{1}{t-1+\alpha_0},
\end{equation}
with the convention that the empty sum is zero.  Hence \(a_{n,n}=0\).  Integral comparison gives, for a constant \(C\) depending only on \(\alpha_0\),
\begin{equation}\label{eq:weight-bound}
0\le a_{n,j}\le C\bigl(1+\log(n/j)\bigr),\qquad 1\le j\le n,
\end{equation}
and \(a_{n,\lceil nx\rceil}\to\log(1/x)\) for every \(x\in(0,1]\).  Since \((1+\log(1/x))^2\) is integrable on \((0,1]\), dominated Riemann-sum convergence yields
\begin{align}
\frac1n\sum_{j=1}^n a_{n,j}&\longrightarrow \int_0^1\log(1/x)\,\dd x=1,\label{eq:weight-mean}\\
\frac1n\sum_{j=1}^n a_{n,j}^2&\longrightarrow \int_0^1\log^2(1/x)\,\dd x=2.\label{eq:weight-second}
\end{align}
Moreover,
\begin{equation}\label{eq:max-weight}
\frac{\max_{1\le j\le n}a_{n,j}}{\sqrt n}\longrightarrow 0.
\end{equation}

\begin{lemma}[Auxiliary concentration and moment bounds]\label{lem:technical}
Let \(p_i>0\) for all \(i\).  All constants below may depend on \(p\), \(m\), and \(\alpha\), but not on \(n\), \(t\), or \(k\).
\begin{enumerate}
\item For every closed set \(F\) in the probability simplex with \(p\notin F\), there are constants \(C,c>0\) such that
\begin{equation}\label{eq:empirical-concentration}
\Pp_p\{\widehat p_k\in F\}\le C e^{-ck},\qquad k\ge1.
\end{equation}
In particular, empirical frequencies leave any fixed compact neighborhood of \(p\) inside the simplex with exponentially small probability.
\item If \(M_k=N(k)-kp\), then for every fixed \(r\ge1\),
\begin{equation}\label{eq:count-moments}
\E_p\norm{M_k}^r=O(k^{r/2}).
\end{equation}
\item If \(\xi_j\) are independent, centered, uniformly bounded real random variables and \(c_{n,j}\) are deterministic weights, then
\begin{equation}\label{eq:weighted-fourth-general}
\E\left|\sum_{j=1}^n c_{n,j}\xi_j\right|^4
\le C_\xi\left[\left(\sum_{j=1}^n c_{n,j}^2\right)^2+
\sum_{j=1}^n c_{n,j}^4\right].
\end{equation}
Consequently, if \(\sum_j c_{n,j}^2=O(n)\) and \(\sum_j c_{n,j}^4=O(n)\), then \(n^{-1/2}\sum_j c_{n,j}\xi_j\) is bounded in \(L^4\).
\end{enumerate}
\end{lemma}

\begin{proof}
The first assertion follows from Hoeffding's inequality applied to the finitely many coordinates, or equivalently from standard finite-alphabet large-deviation bounds.  The moment bound in \eqref{eq:count-moments} follows from the same bounded-summand estimates, for instance by applying the Marcinkiewicz--Zygmund inequality coordinatewise.  Finally, expanding the fourth power of a centered independent sum leaves only pairings and fourth moments, which gives \eqref{eq:weighted-fourth-general}; boundedness of the \(\xi_j\)'s makes the constant uniform.
\end{proof}

\begin{lemma}[Asymptotic linearity]\label{lem:linear}
Under the assumptions of Section~\ref{sec:model},
\begin{align}
L_n^\alpha-nH(p)&=\sum_{j=1}^n\phi(X_j)+O_{L^2}(\log n),\label{eq:linear-L}\\
S_n^\alpha-nH(p)&=\sum_{j=1}^n a_{n,j}\phi(X_j)+O_{L^2}(\log n).\label{eq:linear-S}
\end{align}
\end{lemma}

\begin{proof}
We first consider the log-loss.  The Dirichlet mixture identity gives
\begin{equation}\label{eq:mixture-identity}
\exp(-L_n^\alpha)=\frac{\Gamma(\alpha_0)}{\Gamma(n+\alpha_0)}
\prod_{i=1}^m\frac{\Gamma(N_i(n)+\alpha_i)}{\Gamma(\alpha_i)}.
\end{equation}
Choose \(\delta\in(0,\min_i p_i)\) and let \(A_n=\{\min_i\widehat p_n(i)\ge\delta\}\).  On \(A_n\), Stirling's expansion is uniform over the compact set \(\{r_i\ge\delta,\sum_i r_i=1\}\) and gives
\begin{equation}\label{eq:stirling-L}
L_n^\alpha=nH(\widehat p_n)+\frac{m-1}{2}\log n+O(1).
\end{equation}
By Lemma~\ref{lem:technical}, \(\Pp_p(A_n^c)\le Ce^{-cn}\).  Also \(q_t(X_t)\ge \alpha_{\min}/(t-1+\alpha_0)\), where \(\alpha_{\min}=\min_i\alpha_i\), so \(L_n^\alpha=O(n\log n)\) deterministically.  Therefore the contribution of \(A_n^c\) to the \(L^2\) error is \(O(n^2\log^2n\,e^{-cn})=o(1)\).  Taylor expansion of \(H\) at \(p\) uses that \(\sum_i(\widehat p_n(i)-p_i)=0\), so the constant \(-1\) term in the gradient \(-\log p_i-1\) drops.  On a fixed compact neighborhood of \(p\),
\[
H(\widehat p_n)-H(p)=-\sum_{i=1}^m(\widehat p_n(i)-p_i)\log p_i+O(\norm{\widehat p_n-p}^2).
\]
With \(M_n=N(n)-np\), Lemma~\ref{lem:technical} gives \(\norm{n\norm{\widehat p_n-p}^2}_{L^2}=\norm{\norm{M_n}^2/n}_{L^2}=O(1)\); the exponentially small complement contributes \(o(1)\) in \(L^2\).  Hence
\begin{align}\label{eq:plugin-linear}
n\bigl(H(\widehat p_n)-H(p)\bigr)
&=-\sum_{i=1}^m\bigl(N_i(n)-np_i\bigr)\log p_i
  +O_{L^2}(1) \notag\\
&=\sum_{j=1}^n\phi(X_j)+O_{L^2}(1).
\end{align}
This proves \eqref{eq:linear-L}.

We next consider the predictive self-entropy.  Write \(e_i\) for the \(i\)th coordinate vector and define
\begin{equation}\label{eq:Z-def}
Z_j=e_{X_j}-p,\qquad \E_p Z_j=0.
\end{equation}
Put
\begin{equation}\label{eq:beta-def}
\beta=\alpha-\alpha_0p.
\end{equation}
Then
\begin{equation}\label{eq:qt-minus-p}
q_t-p=\frac{\sum_{j<t}Z_j+\beta}{t-1+\alpha_0}.
\end{equation}
Let \(d_t=q_t-p\) and define the first-order remainder globally by
\begin{equation}\label{eq:R-global}
R_t=H(q_t)-H(p)+\sum_{i=1}^m d_t(i)\log p_i .
\end{equation}
Since \(p\) is in the interior of the simplex, choose \(\eta>0\) so that
\(
\{r: \norm{r-p}_\infty\le \eta,\; \sum_i r_i=1\}
\)
is contained in the interior.  Because \(\sum_i d_t(i)=0\), the constant \(-1\) term in the gradient \(-\log p_i-1\) drops.  On \(G_t=\{\norm{d_t}_\infty\le \eta\}\), Taylor's theorem gives
\begin{equation}\label{eq:first-taylor}
\begin{split}
H(q_t)-H(p)
&=-\sum_i d_t(i)\log p_i+R_t,\\
|R_t|&\le C\norm{d_t}^2\quad\text{on }G_t.
\end{split}
\end{equation}
On \(G_t^c\), the entropy difference and the linear term, and hence \(R_t\), are uniformly bounded by a constant depending only on \(p\), \(m\), and \(\alpha\).  Furthermore, \(\Pp_p(G_t^c)\le C\exp(-ct)\) for all large \(t\), by Lemma~\ref{lem:technical}; the finitely many small \(t\) are absorbed into constants.  Since the centered multinomial count vector \(\sum_{j<t}Z_j\) has fourth moments of order \(t^2\), again by Lemma~\ref{lem:technical},
\begin{equation}\label{eq:dt-L4}
\norm{d_t}_{L^4}=O(t^{-1/2}).
\end{equation}
Consequently,
\begin{equation}\label{eq:R-L2}
\norm{R_t}_{L^2}\le C\norm{d_t}_{L^4}^2+C\Pp_p(G_t^c)^{1/2}=O(t^{-1}),
\end{equation}
with the finitely many small values of \(t\) absorbed into the constant.  Minkowski's inequality gives
\begin{equation}\label{eq:R-sum-L2}
\left\|\sum_{t=1}^nR_t\right\|_{L^2}
\le \sum_{t=1}^n\norm{R_t}_{L^2}=O(\log n).
\end{equation}
Thus the first-order Taylor remainder is controlled in \(L^2\).

It remains to identify the linear part.  Its random component equals
\begin{align}
-&\sum_{t=1}^n\sum_{i=1}^m
\frac{\sum_{j<t}Z_j(i)}{t-1+\alpha_0}\log p_i \notag\\
&=-\sum_{j=1}^n\sum_{i=1}^m Z_j(i)\log p_i
\sum_{t=j+1}^n\frac{1}{t-1+\alpha_0}\label{eq:linear-reorder}\\
&=\sum_{j=1}^n a_{n,j}\phi(X_j),\label{eq:linear-identify}
\end{align}
because \(-\sum_iZ_j(i)\log p_i=-\log p_{X_j}-H(p)=\phi(X_j)\).  The deterministic prior-drift component is
\begin{equation}\label{eq:prior-drift}
-\sum_{t=1}^n\sum_{i=1}^m\frac{\beta_i}{t-1+\alpha_0}\log p_i=O(\log n).
\end{equation}
Combining \eqref{eq:first-taylor}--\eqref{eq:prior-drift} proves \eqref{eq:linear-S}.
\end{proof}

\section{Proofs of the Main Results}\label{sec:proofs}
\begin{proof}[Proof of Theorem~\ref{thm:main}]
By Lemma~\ref{lem:linear}, it suffices first to prove a joint central limit theorem for
\begin{equation}\label{eq:array-vector}
\frac1{\sqrt n}\left(\sum_{j=1}^n\phi(X_j),\sum_{j=1}^n a_{n,j}\phi(X_j)\right).
\end{equation}
This is a triangular array of independent centered bounded random vectors.  Its covariance matrix is
\begin{equation}\label{eq:array-cov}
V(p)
\begin{pmatrix}
1& n^{-1}\sum_{j=1}^n a_{n,j}\\[1mm]
n^{-1}\sum_{j=1}^n a_{n,j}&n^{-1}\sum_{j=1}^n a_{n,j}^2
\end{pmatrix},
\end{equation}
which converges to the matrix in \eqref{eq:main-clt} by \eqref{eq:weight-mean} and \eqref{eq:weight-second}.  The Lindeberg condition follows from boundedness of \(\phi\) and from \eqref{eq:max-weight}.  The Cram\'{e}r-Wold device gives the weak convergence of \eqref{eq:array-vector}.  The \(O_{L^2}(\log n)\) remainders in Lemma~\ref{lem:linear} are negligible after division by \(\sqrt n\), proving \eqref{eq:main-clt}.

For convergence of the second moments, note that \(\phi\) is bounded and the weight estimates imply \(\sum_j a_{n,j}^2=O(n)\) and \(\sum_j a_{n,j}^4=O(n)\).  Lemma~\ref{lem:technical} yields, uniformly in \(n\),
\begin{equation}\label{eq:fourth-moment}
\E_p\left|\frac1{\sqrt n}\sum_{j=1}^n c_{n,j}\phi(X_j)\right|^4=O(1)
\end{equation}
for \(c_{n,j}=1\), for \(c_{n,j}=a_{n,j}\), and for each fixed linear combination of these two choices.  Hence the squared norm of the triangular-array vector is uniformly integrable.  Together with the \(L^2\)-negligible remainders, this proves \eqref{eq:mse-L} and \eqref{eq:mse-S}.  Since \(V(p)>0\), the ratio in \eqref{eq:mse-ratio} converges to two.
\end{proof}

\begin{proof}[Proof of Corollary~\ref{cor:corrections}]
The assumptions imply \(B_n/\sqrt n\to0\) and \(C_n/\sqrt n\to0\) in \(L^2\), hence in probability.  Slutsky's theorem gives the same weak limit as in Theorem~\ref{thm:main}.  The same \(L^2\) negligibility gives the unchanged leading mean-square-error constants.
\end{proof}

\begin{proof}[Proof of Proposition~\ref{prop:bias}]
The first assertion follows from \eqref{eq:mixture-identity} and Stirling's expansion, equivalently from \eqref{eq:stirling-L} together with \(\E_p[H(\widehat p_n)-H(p)]=O(n^{-1})\).

We prove \eqref{eq:bias-S}.  Put \(k=t-1\), \(M_k=N(k)-kp\), and \(\beta=\alpha-\alpha_0p\), so that
\begin{equation}\label{eq:q-k-bias-proof}
q_{k+1}-p=\frac{M_k+\beta}{k+\alpha_0}.
\end{equation}
Let \(u_k=q_{k+1}-p\) and define the second-order remainder globally by
\begin{equation}\label{eq:rho-global}
\rho_k=H(p+u_k)-H(p)+\sum_{i=1}^m u_k(i)\log p_i
+\frac12\sum_{i=1}^m\frac{u_k(i)^2}{p_i}.
\end{equation}
Choose \(\eta>0\) so that the \(\ell_\infty\)-ball of radius \(\eta\) around \(p\), intersected with the simplex, is contained in the interior.  For \(k\) below a fixed threshold, all contributions to \(\sum_k \E_p[H(p+u_k)-H(p)]\) are bounded and hence amount to \(O(1)\).  For larger \(k\), set \(G_k=\{\norm{u_k}_\infty\le\eta\}\).  Lemma~\ref{lem:technical} and the deterministic drift \(\beta/(k+\alpha_0)\) imply \(\Pp_p(G_k^c)\le Ce^{-ck}\).  Since \(\sum_i u_k(i)=0\), the constant \(-1\) term in the gradient of entropy again drops, and the Hessian on the simplex has diagonal entries \(-1/p_i\).  On \(G_k\), Taylor's theorem gives
\begin{equation}\label{eq:bias-taylor}
H(p+u_k)-H(p)=
-\sum_{i=1}^m u_k(i)\log p_i
-\frac12\sum_{i=1}^m\frac{u_k(i)^2}{p_i}
+\rho_k.
\end{equation}
The remainder satisfies \(|\rho_k|\le C\norm{u_k}^3\) on \(G_k\); on \(G_k^c\), the entropy difference and the displayed linear and quadratic terms, and hence \(\rho_k\), are bounded by constants depending only on \(p\), \(m\), and \(\alpha\).  Since \(\E_p\norm{M_k}^3=O(k^{3/2})\), we have \(\E_p|\rho_k|=O(k^{-3/2})+O(e^{-ck})\) for large \(k\).  Therefore \(\sum_{k=0}^{n-1}\E_p\rho_k=O(1)\).

The expectation of the linear term is
\begin{equation}\label{eq:bias-linear}
-\sum_{k=0}^{n-1}\sum_{i=1}^m\frac{\beta_i}{k+\alpha_0}\log p_i
=-\left(\sum_{i=1}^m\beta_i\log p_i\right)\log n+O(1).
\end{equation}
For the quadratic term, \(\E_pM_k(i)=0\) and \(\E_pM_k(i)^2=kp_i(1-p_i)\), so
\begin{equation}\label{eq:bias-quadratic-moment}
\sum_{i=1}^m\frac{\E_p(M_k(i)+\beta_i)^2}{p_i}
=k(m-1)+\sum_{i=1}^m\frac{\beta_i^2}{p_i}.
\end{equation}
Consequently,
\begin{align}\label{eq:bias-quadratic}
&-\frac12\sum_{k=0}^{n-1}\frac{1}{(k+\alpha_0)^2}
\left(k(m-1)+\sum_{i=1}^m\frac{\beta_i^2}{p_i}\right) \notag\\
&\qquad=-\frac{m-1}{2}\log n+O(1).
\end{align}
Adding \eqref{eq:bias-linear}, \eqref{eq:bias-quadratic}, and the remainder contribution proves \eqref{eq:bias-S} and \eqref{eq:bias-coeff}.
\end{proof}

\section{Finite-Order Markov Sources}\label{sec:markov}
The i.i.d.\ theorem has a fixed finite-order Markov analogue, but not by simply replacing \(V(p)\) in Theorem~\ref{thm:main} by the asymptotic variance of the Markov information density.  The Markov variance separates context-entropy fluctuations, which enter both quantities without logarithmic reweighting, from within-context information innovations, which are reused by later predictions as in the i.i.d.\ proof.

Fix an order \(r\ge1\).  Let \(W_t=X_{t-r}^{t-1}\in\Wcal=\Acal^r\) be the context before \(X_t\), for \(t\ge r+1\).  Assume that \((X_t)\) is stationary and that
\begin{equation}\label{eq:markov-transition}
P_w(a)=\Pp\{X_t=a\mid W_t=w\}>0,
        \qquad w\in\Wcal,\; a\in\Acal.
\end{equation}
Then the context chain \((W_t)\) is a finite irreducible aperiodic Markov chain.  Its stationary distribution is denoted by \(\pi\).  All probabilities and expectations in this section are taken under this stationary Markov source.  Put
\begin{equation}\label{eq:markov-Hw}
H_w=-\sum_{a\in\Acal}P_w(a)\log P_w(a),
        \qquad
h=\sum_{w\in\Wcal}\pi(w)H_w.
\end{equation}
For each context \(w\), fix strictly positive pseudo-counts \(\alpha_w=(\alpha_{w,a})_{a\in\Acal}\) and set \(\alpha_{w,0}=\sum_a\alpha_{w,a}\).  Let
\begin{align}
N_t(w,a)&=\sum_{s=r+1}^t \ind\{W_s=w,\;X_s=a\},\label{eq:markov-count-wa}\\
N_t(w)&=\sum_{a\in\Acal}N_t(w,a).\label{eq:markov-count-w}
\end{align}
The context-wise Dirichlet predictor is
\begin{equation}\label{eq:markov-predictor}
q_t(a)=\frac{N_{t-1}(W_t,a)+\alpha_{W_t,a}}
             {N_{t-1}(W_t)+\alpha_{W_t,0}},
        \qquad t\ge r+1.
\end{equation}
As in the i.i.d.\ case, \(H(q_t)\) can be updated online by maintaining, for each context \(w\), the sum \(\sum_a c_{w,a}\log c_{w,a}\).  Only the active context \(W_t\) and the observed transition symbol are updated, so the per-symbol update is \(O(1)\) for fixed \(m\) and \(r\).

Define
\begin{equation}\label{eq:markov-LS}
L_{n,r}^{\alpha}=-\sum_{t=r+1}^n\log q_t(X_t),
        \qquad
S_{n,r}^{\alpha}=\sum_{t=r+1}^n H(q_t).
\end{equation}
The missing first \(r\) symbols are boundary terms and are immaterial at the \(\sqrt n\) scale.

Let
\begin{equation}\label{eq:markov-I-D-eps}
I_t=-\log P_{W_t}(X_t),\qquad
D_t=H_{W_t}-h,
        \qquad
\varepsilon_t=I_t-H_{W_t}.
\end{equation}
Then \(I_t-h=D_t+\varepsilon_t\) and \(\E[\varepsilon_t\mid W_t]=0\).  Define the mean conditional varentropy
\begin{equation}\label{eq:markov-nu}
\nu=\E\varepsilon_t^2
=\sum_{w\in\Wcal}\pi(w)\sum_{a\in\Acal}P_w(a)
\bigl(-\log P_w(a)-H_w\bigr)^2.
\end{equation}
Let \(P\) also denote the transition matrix of the context chain.  For \(f(w)=H_w-h\), let \(g\) be the bounded solution of the Poisson equation
\begin{equation}\label{eq:poisson}
g(w)-(Pg)(w)=f(w),
        \qquad \sum_{w\in\Wcal}\pi(w)g(w)=0.
\end{equation}
Set
\begin{equation}\label{eq:markov-M}
M_t=g(W_{t+1})-(Pg)(W_t)
\end{equation}
and
\begin{equation}\label{eq:markov-sigma}
\sigma^2=\E\bigl(M_t+\varepsilon_t\bigr)^2.
\end{equation}
Equivalently, \(\sigma^2\) is the asymptotic variance of the true information-density process:
\begin{equation}\label{eq:sigma-info-density}
\sigma^2=\lim_{n\to\infty}\frac1n
\Var\left(\sum_{t=r+1}^n(I_t-h)\right).
\end{equation}

\begin{theorem}[Finite-order Markov CLT]\label{thm:markov}
Under the assumptions above,
\begin{equation}\label{eq:markov-clt}
\sqrt n\left(\frac{L_{n,r}^{\alpha}}{n}-h,
             \frac{S_{n,r}^{\alpha}}{n}-h\right)
\Rightarrow
\Normal\left(0,
\begin{pmatrix}
\sigma^2&\sigma^2\\[1mm]
\sigma^2&\sigma^2+\nu
\end{pmatrix}\right).
\end{equation}
Moreover,
\begin{align}
 n\E\left[\left(\frac{L_{n,r}^{\alpha}}{n}-h\right)^2\right]
        &\longrightarrow \sigma^2,\label{eq:markov-mse-L}\\
 n\E\left[\left(\frac{S_{n,r}^{\alpha}}{n}-h\right)^2\right]
        &\longrightarrow \sigma^2+\nu.\label{eq:markov-mse-S}
\end{align}
If \(\sigma^2>0\), then
\begin{equation}\label{eq:markov-mse-ratio}
\frac{\E[(S_{n,r}^{\alpha}/n-h)^2]}
     {\E[(L_{n,r}^{\alpha}/n-h)^2]}
\longrightarrow 1+\frac{\nu}{\sigma^2}.
\end{equation}
\end{theorem}

The proof uses two standard finite-state estimates.  They are recorded explicitly because they are the only additional ingredients beyond the i.i.d.\ argument.

\begin{lemma}[Finite-state weighted estimates]\label{lem:markov-estimates}
Under \eqref{eq:markov-transition}, the following bounds hold for constants depending only on the transition matrix and on fixed moments.
\begin{enumerate}
\item For each \(w\in\Wcal\), \(a\in\Acal\), and each fixed \(q\ge1\),
\begin{align}
\E|N_t(w)-\pi(w)t|^q&=O(t^{q/2}),\label{eq:markov-count-moment}\\
\E\left\|\sum_{s=r+1}^t \ind\{W_s=w\}(e_{X_s}-P_w)\right\|^q&=O(t^{q/2}).\label{eq:markov-transition-moment}
\end{align}
Moreover, empirical context and transition frequencies have exponentially small deviations from their stationary limits.
\item Let \(F\) be a bounded real function on \(\Wcal\), and let \(c_{n,t}\), \(r+1\le t\le n\), be deterministic weights with \(n^{-1}\sum_t c_{n,t}^2=O(1)\).  Put \(\bar F=\sum_w\pi(w)F(w)\).  Then
\begin{equation}\label{eq:weighted-ergodic-general}
\frac1n\sum_{t=r+1}^n c_{n,t}\{F(W_t)-\bar F\}
\longrightarrow0
\end{equation}
in \(L^2\).  In particular, for
\begin{equation}\label{eq:markov-det-weight}
b_{n,t}=\sum_{u=t+1}^n\frac1u,
        \qquad r+1\le t\le n,
\end{equation}
one has
\begin{align}
\frac1n\sum_{t=r+1}^n b_{n,t}F(W_t)
&\longrightarrow \sum_w\pi(w)F(w),\label{eq:weighted-ergodic-1}\\
\frac1n\sum_{t=r+1}^n b_{n,t}^2F(W_t)
&\longrightarrow 2\sum_w\pi(w)F(w)\label{eq:weighted-ergodic-2}
\end{align}
in \(L^2\).
\end{enumerate}
\end{lemma}

\begin{proof}
For the finite irreducible aperiodic context chain induced by \eqref{eq:markov-transition}, the transition matrix has a spectral gap on mean-zero functions.  This gives exponential covariance decay, and Chernoff--Hoeffding type bounds for finite Markov chains give exponential concentration for bounded additive functionals \cite{Lezaud1998}.  Equivalently, these are finite-state specializations of geometric-ergodic Markov-chain estimates \cite{MeynTweedie2009}.  Integrating the resulting tails gives the square-root moment growth in \eqref{eq:markov-count-moment}.  The second bound in \eqref{eq:markov-transition-moment} is also obtained directly from martingale moment inequalities, since \(\ind\{W_s=w\}(e_{X_s}-P_w)\) is a bounded martingale difference with respect to the natural filtration.

To prove the weighted ergodic bound, use stationarity and exponential mixing:
\(|\Cov(F(W_s),F(W_t))|\le C\rho^{|t-s|}\) for some \(\rho<1\).  The variance of the left side of \eqref{eq:weighted-ergodic-general} is bounded by
\begin{equation*}
\frac{C}{n^2}\sum_{s,t=r+1}^n |c_{n,s}c_{n,t}|\rho^{|s-t|}
\le \frac{C'}{n^2}\sum_{t=r+1}^n c_{n,t}^2=o(1).
\end{equation*}
The displayed limits for \(b_{n,t}\) follow from Riemann-sum convergence.  For \(x\in(0,1]\),
\begin{equation*}
 b_{n,\lceil nx\rceil}\to\log(1/x),\quad
 \frac1n\sum_t b_{n,t}\to1,\quad
 \frac1n\sum_t b_{n,t}^2\to2,
\end{equation*}
and \(n^{-1}\sum_t b_{n,t}^4=O(1)\).
\end{proof}

\begin{lemma}[Replacement of random context-reuse weights]\label{lem:markov-weight-replacement}
Let
\begin{equation}\label{eq:markov-exact-weight}
B_{n,t}=\sum_{u=t+1}^n
\frac{\ind\{W_u=W_t\}}{N_{u-1}(W_t)+\alpha_{W_t,0}},
        \qquad r+1\le t\le n.
\end{equation}
Then
\begin{equation}\label{eq:markov-weight-replacement}
\frac1{\sqrt n}\sum_{t=r+1}^n(B_{n,t}-b_{n,t})\varepsilon_t
\longrightarrow0
\end{equation}
in \(L^2\), where \(b_{n,t}\) is defined in \eqref{eq:markov-det-weight}.
\end{lemma}

\begin{proof}
Let \(K_t=N_t(W_t)\) be the visit rank of the active context at time \(t\); this is known before \(X_t\) is observed.  If \(W_t=w\), then future visits to the same context have previous-visit counts \(K_t,K_t+1,\ldots,N_n(w)-1\), and hence
\begin{equation}\label{eq:B-rank-form}
B_{n,t}=\sum_{\ell=K_t}^{N_n(w)-1}\frac1{\ell+\alpha_{w,0}}.
\end{equation}
The elementary harmonic estimate
\begin{equation}\label{eq:harmonic-estimate}
\left|\sum_{\ell=K}^{N-1}\frac1{\ell+\alpha}-\log\frac{N}{K}\right|
\le \frac{C}{K},\qquad 1\le K\le N,
\end{equation}
with the convention that the empty sum is zero, and the corresponding estimate
\(|b_{n,t}-\log(n/t)|\le C/t\), give
\begin{equation}\label{eq:B-b-decomp}
B_{n,t}-b_{n,t}=A_{n,W_t}-C_t+R_{n,t},
\end{equation}
where
\begin{equation*}
A_{n,w}=\log\frac{N_n(w)}{\pi(w)n},\qquad
C_t=\log\frac{K_t}{\pi(W_t)t},
\end{equation*}
on the event \(N_n(w)>0\), with harmless bounded definitions on its exponentially small complement.  The remainder satisfies, on the same good event,
\begin{equation}\label{eq:R-bound-clean}
|R_{n,t}|\le C\bigl(K_t^{-1}+t^{-1}\bigr).
\end{equation}
The complement of the good event is exponentially small; its contribution to all bounds below is negligible because \(|B_{n,t}|+|b_{n,t}|\le C(1+\log n)\) and \(\varepsilon_t\) is bounded.

The concentration and moment estimates in Lemma~\ref{lem:markov-estimates} imply
\begin{equation}\label{eq:A-C-moments}
\E |A_{n,w}|^4=O(n^{-2}),\qquad
\E |C_t|^2=O(t^{-1}).
\end{equation}
Indeed, on the good event the logarithms are bounded by a constant multiple of the relative count error, while the bad event is exponentially small.

We treat the three terms in \eqref{eq:B-b-decomp}.  For the final-count term, put
\(E_{n,w}=\sum_{t=r+1}^n\ind\{W_t=w\}\varepsilon_t\).  This is a bounded martingale sum with \(\E E_{n,w}^4=O(n^2)\).  Hence, by Cauchy's inequality,
\begin{equation}\label{eq:final-count-term}
\left\|\sum_w A_{n,w}E_{n,w}\right\|_2=O(1).
\end{equation}
For the rank term, \(C_t\) is predictable and \(\E[\varepsilon_t\mid\mathcal F_{t-1}]=0\).  Orthogonality of martingale differences gives
\begin{align}
\left\|\sum_{t=r+1}^n C_t\varepsilon_t\right\|_2^2
&=\sum_{t=r+1}^n\E[C_t^2\varepsilon_t^2]\notag\\
&=O\left(\sum_{t=1}^n t^{-1}\right)=O(\log n).\label{eq:rank-term}
\end{align}
For the remainder we use only the deterministic bound \eqref{eq:R-bound-clean}, avoiding any orthogonality assertion for \(R_{n,t}\).  For each context \(w\),
\begin{equation*}
\sum_{t:W_t=w}\frac1{K_t}
=\sum_{k=1}^{N_n(w)}\frac1k
=O(\log n),
\end{equation*}
and \(\sum_{t=1}^n t^{-1}=O(\log n)\).  Since \(\varepsilon_t\) is bounded, the good-event contribution satisfies
\begin{align}
\left|\sum_{t=r+1}^n R_{n,t}\varepsilon_t\right|
&\le C\sum_{t=r+1}^n\bigl(K_t^{-1}+t^{-1}\bigr) \notag\\
&=O(\log n).\label{eq:remainder-term}
\end{align}
The exponentially small complement contributes \(o(1)\) in \(L^2\) after the crude bound \(O(n\log n)\).  Together, the final-count, rank, and remainder bounds give an unnormalized \(L^2\) bound \(O(\log n)\), which is \(o(\sqrt n)\).  This proves \eqref{eq:markov-weight-replacement}.
\end{proof}

\begin{proof}[Proof of Theorem~\ref{thm:markov}]
We write the proof for stationary initialization.  A fixed non-stationary initial context changes only boundary and exponentially decaying transient terms for the finite irreducible aperiodic context chain.

First, the Dirichlet mixture identity factorizes over contexts.  Uniform Stirling expansion on the event that all empirical transition probabilities remain in a fixed compact neighborhood of the true transition probabilities gives
\begin{equation}\label{eq:markov-stirling}
L_{n,r}^{\alpha}=\sum_{w\in\Wcal}N_n(w)H(\widehat P_w)
        +\frac{|\Wcal|(m-1)}{2}\log n+O(1),
\end{equation}
where \(\widehat P_w(a)=N_n(w,a)/N_n(w)\).  The exceptional event has exponentially small probability by Lemma~\ref{lem:markov-estimates}, and \(L_{n,r}^{\alpha}=O(n\log n)\) deterministically because all pseudo-counts are positive.  Its contribution is therefore negligible in \(L^2\).  Put \(\Delta_n(w,a)=N_n(w,a)-N_n(w)P_w(a)\).  Taylor expansion of \(H\) at each \(P_w\), again using the simplex constraint to remove the constant \(-1\) term in the entropy gradient, yields
\begin{align}
&\sum_{w\in\Wcal}N_n(w)\{H(\widehat P_w)-H_w\} \notag\\
&= -\sum_{w\in\Wcal}\sum_{a\in\Acal}\Delta_n(w,a)\log P_w(a)
   +O_{L^2}(1) \notag\\
&= \sum_{t=r+1}^n \varepsilon_t+O_{L^2}(1),\label{eq:markov-plugin-linear}
\end{align}
The \(O_{L^2}(1)\) remainder follows from \eqref{eq:markov-transition-moment}: on the good event it is bounded by a constant times \(\sum_w N_n(w)\|\widehat P_w-P_w\|^2\), and the fourth moments of the corresponding centered transition counts are \(O(n^2)\).  Since \(\sum_wN_n(w)H_w-nh=\sum_{t=r+1}^nD_t+O(1)\),
\begin{equation}\label{eq:markov-linear-L}
\begin{split}
L_{n,r}^{\alpha}-nh
&=\sum_{t=r+1}^n(D_t+\varepsilon_t)\\
&\quad+O_{L^2}(\log n).
\end{split}
\end{equation}

For the predictive self-entropy, put
\begin{equation}\label{eq:markov-beta}
\beta_w=\alpha_w-\alpha_{w,0}P_w.
\end{equation}
If \(W_t=w\), then
\begin{equation}\label{eq:markov-qt-minus}
q_t-P_w=\frac{\sum_{s<t}\ind\{W_s=w\}(e_{X_s}-P_w)+\beta_w}
               {N_{t-1}(w)+\alpha_{w,0}}.
\end{equation}
Define the first-order Taylor remainder \(R_t\) by
\begin{equation}\label{eq:markov-R-def}
R_t=H(q_t)-H_{W_t}
+\sum_{a\in\Acal}\bigl(q_t(a)-P_{W_t}(a)\bigr)\log P_{W_t}(a).
\end{equation}
For all sufficiently large \(t\), on the event \(N_{t-1}(W_t)\ge c t\) and \(q_t\) is in a fixed interior neighborhood of \(P_{W_t}\), Taylor's theorem gives \(|R_t|\le C\|q_t-P_{W_t}\|^2\).  The complement has exponentially small probability by Lemma~\ref{lem:markov-estimates}; the finitely many smaller \(t\) are absorbed into constants.  From \eqref{eq:markov-transition-moment} and \eqref{eq:markov-qt-minus}, \(\|q_t-P_{W_t}\|_{L^4}=O(t^{-1/2})\).  Therefore
\begin{equation}\label{eq:markov-R-bound}
\|R_t\|_{L^2}=O(t^{-1}),
        \qquad
\left\|\sum_{t=r+1}^nR_t\right\|_{L^2}=O(\log n).
\end{equation}
Reordering the random linear term gives the exact context-reuse weight \eqref{eq:markov-exact-weight}; the deterministic prior-drift terms contribute only \(O(\log n)\).  Hence
\begin{equation}\label{eq:markov-linear-S-exact}
S_{n,r}^{\alpha}-nh=\sum_{t=r+1}^nD_t+
\sum_{t=r+1}^nB_{n,t}\varepsilon_t+O_{L^2}(\log n).
\end{equation}
By Lemma~\ref{lem:markov-weight-replacement},
\begin{equation}\label{eq:markov-linear-S}
S_{n,r}^{\alpha}-nh=\sum_{t=r+1}^nD_t+
\sum_{t=r+1}^n b_{n,t}\varepsilon_t+o_{L^2}(\sqrt n).
\end{equation}

The context-entropy term is handled by the Poisson equation.  From \eqref{eq:poisson} and \eqref{eq:markov-M},
\begin{equation}\label{eq:poisson-decomp}
D_t=g(W_t)-g(W_{t+1})+M_t.
\end{equation}
Therefore \(\sum_tD_t=\sum_tM_t+O(1)\).  Equations \eqref{eq:markov-linear-L} and \eqref{eq:markov-linear-S} reduce the problem to the triangular martingale array
\begin{equation}\label{eq:markov-array}
\frac1{\sqrt n}\sum_{t=r+1}^n
\begin{pmatrix}
M_t+\varepsilon_t\\[1mm]
M_t+b_{n,t}\varepsilon_t
\end{pmatrix}.
\end{equation}
With respect to the natural filtration \(\mathcal F_t=\sigma(X_1^t)\), the summands are martingale differences.  They are bounded up to the deterministic factor \(b_{n,t}\), and \(\max_t b_{n,t}/\sqrt n\to0\); hence Lindeberg's condition holds.

For the conditional covariance, define bounded functions of the context
\begin{align}
A(w)&=\E[M_t^2\mid W_t=w],\notag\\
B(w)&=\E[M_t\varepsilon_t\mid W_t=w],\notag\\
C(w)&=\E[\varepsilon_t^2\mid W_t=w].\label{eq:ABC}
\end{align}
Lemma~\ref{lem:markov-estimates} gives the limits of the weighted ergodic averages in the conditional covariance sums.  Since \(n^{-1}\sum_t b_{n,t}\to1\) and \(n^{-1}\sum_t b_{n,t}^2\to2\), the weighted cross and lower-right terms have the required limits.  Consequently the conditional covariance matrix of \eqref{eq:markov-array} converges in probability to
\begin{equation}\label{eq:markov-cov-limit}
\begin{pmatrix}
\E[A+2B+C]&\E[A+2B+C]\\[1mm]
\E[A+2B+C]&\E[A+2B+2C]
\end{pmatrix}
=
\begin{pmatrix}
\sigma^2&\sigma^2\\[1mm]
\sigma^2&\sigma^2+\nu
\end{pmatrix}.
\end{equation}
The martingale central limit theorem proves \eqref{eq:markov-clt}; see, for example, \cite{HallHeyde1980}.

Finally, \(\sum_tb_{n,t}^2=O(n)\) and \(\sum_tb_{n,t}^4=O(n)\).  Burkholder's inequality for martingales therefore gives uniform \(L^4\) bounds for the normalized sums in \eqref{eq:markov-array}.  Together with the negligible remainders in \eqref{eq:markov-linear-L} and \eqref{eq:markov-linear-S}, this gives uniform integrability of the squared normalized errors and proves \eqref{eq:markov-mse-L}--\eqref{eq:markov-mse-S}.  The ratio \eqref{eq:markov-mse-ratio} follows when \(\sigma^2>0\).
\end{proof}

In the formal one-context, order-zero specialization corresponding to the memoryless case, \(D_t=0\), \(M_t=0\), and \(\nu=\sigma^2=V(p)\); the Markov covariance matrix therefore reduces to the covariance matrix in Theorem~\ref{thm:main}.  More generally, the i.i.d.\ factor-two variance ratio persists whenever the context-entropy component is absent, for example when \(H_w\) is constant over recurrent contexts and hence \(D_t\equiv0\); otherwise the ratio is \(1+\nu/\sigma^2\), not necessarily two.

\section{Interpretation}\label{sec:interpretation}

The factor two in Theorem~\ref{thm:main} is a consequence of how
past observations are reused by the sequential predictor.  The realized
code length is driven, to first order, by the ordinary empirical sum of
the information variables \(-\log p_{X_j}\).  Predictive self-entropy is
different: observation \(X_j\) affects all later predictive distributions
and therefore enters the linearization with the reuse weight
\(a_{n,j}\sim \log(n/j)\).  These weights have average one but second
moment two on the unit interval.  Thus the two quantities share the same
first-order entropy target, while their central-limit fluctuations differ.

The Markov theorem identifies which part of this mechanism survives in
the presence of contexts.  Within each recurrent context, conditional
information innovations are reused through the same logarithmic kernel
and therefore receive the additional variance contribution.  Fluctuations
of the visited contexts themselves behave differently: they enter through
the active-context entropies and appear in realized log-loss and predictive
self-entropy with the same weight.  The variance ratio is therefore
\(1+\nu/\sigma^2\), with the i.i.d.\ factor two recovered precisely when
the context-entropy component is absent.

This distinction is not an efficiency claim against predictive
self-entropy.  Rather, it shows that a natural forecast-only quantity---the
self-expected logarithmic score of each issued forecast---has a different
stochastic linearization from the realized prequential log-loss.  The two
normalized quantities estimate the same entropy or entropy rate to first
order, but their second-order variance and bias structures are not the
same.

The bias calculation reinforces this separation.  The logarithmic
coefficient of \(L_n^\alpha\) is the usual parametric redundancy
coefficient \((m-1)/2\), whereas the coefficient of \(S_n^\alpha\) in
\eqref{eq:bias-coeff} depends on both the source distribution \(p\) and
the pseudo-count vector \(\alpha\).  Although such logarithmic terms can
be estimated or subtracted, Corollary~\ref{cor:corrections} shows that
they do not affect the leading \(1/n\) mean-square-error constants when
\(V(p)>0\).

\section{Conclusion and Outlook}\label{sec:conclusion}

For Dirichlet universal coding, realized arithmetic code length and
cumulative predictive self-entropy are first-order equivalent but
second-order distinct.  In the memoryless case this distinction produces
the factor-two variance and mean-square-error constants.  In the
fixed-order Markov case, the same reuse mechanism acts within contexts,
while context-entropy fluctuations contribute equally to both quantities;
this yields the context/innovation covariance decomposition of
Theorem~\ref{thm:markov}.

A natural next step is to extend the analysis to finite variable-memory
sources and context-tree predictors.  If the relevant context tree is
known, the active context is a finite-state function of the past and the
argument should reduce to a finite-state version of
Theorem~\ref{thm:markov}.  For Bayesian context-tree or CTW predictors
with an unknown tree, one would additionally need an oracle-equivalence
result showing that the mixture predictor differs from the true-tree
predictor by only logarithmic order, both for realized log-loss and for
cumulative predictive self-entropy.  Proving such an oracle step is a
separate model-selection problem and is left for future work.

\section*{Statements and Declarations}

\smallskip
\noindent\textbf{Funding and competing interests.}
No funding or other support was received for this work. The author
declares no competing interests.

\smallskip
\noindent\textbf{Data availability.}
This work is theoretical and does not use empirical datasets.

\smallskip
\noindent\textbf{Use of generative artificial intelligence.}
Generative AI assisted with language and structural editing,
literature searches, and consistency checks. The author
assumes full responsibility for the mathematical content, references,
and final manuscript.


\begin{thebibliography}{99}

\bibitem{Dawid1984}
A. P. Dawid,
``Present position and potential developments: Some personal views: Statistical theory: The prequential approach,''
\emph{Journal of the Royal Statistical Society: Series A (General)}, vol. 147, no. 2, pp. 278--292, 1984,
doi: 10.2307/2981683.

\bibitem{Dawid1999}
A. P. Dawid,
``Prequential probability: Principles and properties,''
\emph{Bernoulli}, vol. 5, no. 1, pp. 125--162, Feb. 1999,
doi: 10.2307/3318616.

\bibitem{KrichevskyTrofimov1981}
R. E. Krichevsky and V. K. Trofimov,
``The performance of universal encoding,''
\emph{IEEE Transactions on Information Theory}, vol. IT-27, no. 2, pp. 199--207, Mar. 1981,
doi: 10.1109/TIT.1981.1056331.

\bibitem{Rissanen1984}
J. Rissanen,
``Universal coding, information, prediction, and estimation,''
\emph{IEEE Transactions on Information Theory}, vol. 30, no. 4, pp. 629--636, Jul. 1984,
doi: 10.1109/TIT.1984.1056936.

\bibitem{ClarkeBarron1990}
B. S. Clarke and A. R. Barron,
``Information-theoretic asymptotics of Bayes methods,''
\emph{IEEE Transactions on Information Theory}, vol. 36, no. 3, pp. 453--471, May 1990,
doi: 10.1109/18.54897.

\bibitem{MerhavFeder1998}
N. Merhav and M. Feder,
``Universal prediction,''
\emph{IEEE Transactions on Information Theory}, vol. 44, no. 6, pp. 2124--2147, Oct. 1998,
doi: 10.1109/18.720534.

\bibitem{XieBarron2000}
Q. Xie and A. R. Barron,
``Asymptotic minimax regret for data compression, gambling, and prediction,''
\emph{IEEE Transactions on Information Theory}, vol. 46, no. 2, pp. 431--445, Mar. 2000,
doi: 10.1109/18.825803.

\bibitem{Good1952}
I. J. Good,
``Rational decisions,''
\emph{Journal of the Royal Statistical Society: Series B (Methodological)}, vol. 14, no. 1, pp. 107--114, 1952,
doi: 10.1111/j.2517-6161.1952.tb00104.x.

\bibitem{Savage1971}
L. J. Savage,
``Elicitation of personal probabilities and expectations,''
\emph{Journal of the American Statistical Association}, vol. 66, no. 336, pp. 783--801, Dec. 1971,
doi: 10.1080/01621459.1971.10482346.

\bibitem{GneitingRaftery2007}
T. Gneiting and A. E. Raftery,
``Strictly proper scoring rules, prediction, and estimation,''
\emph{Journal of the American Statistical Association}, vol. 102, no. 477, pp. 359--378, Mar. 2007,
doi: 10.1198/016214506000001437.

\bibitem{GrunwaldDawid2004}
P. D. Gr\"unwald and A. P. Dawid,
``Game theory, maximum entropy, minimum discrepancy and robust Bayesian decision theory,''
\emph{The Annals of Statistics}, vol. 32, no. 4, pp. 1367--1433, Aug. 2004,
doi: 10.1214/009053604000000553.

\bibitem{Miller1955}
G. A. Miller,
``Note on the bias of information estimates,''
in \emph{Information Theory in Psychology: Problems and Methods}, H. Quastler, Ed.
Glencoe, IL, USA: Free Press, 1955, pp. 95--100.

\bibitem{Basharin1959}
G. P. Basharin,
``On a statistical estimate for the entropy of a sequence of independent random variables,''
\emph{Theory of Probability and Its Applications}, vol. 4, no. 3, pp. 333--336, 1959,
doi: 10.1137/1104033.

\bibitem{AntosKontoyiannis2001}
A. Antos and I. Kontoyiannis,
``Convergence properties of functional estimates for discrete distributions,''
\emph{Random Structures \& Algorithms}, vol. 19, no. 3--4, pp. 163--193, 2001,
doi: 10.1002/rsa.10019.

\bibitem{Paninski2003}
L. Paninski,
``Estimation of entropy and mutual information,''
\emph{Neural Computation}, vol. 15, no. 6, pp. 1191--1253, Jun. 2003,
doi: 10.1162/089976603321780272.

\bibitem{WolpertWolf1995}
D. H. Wolpert and D. R. Wolf,
``Estimating functions of probability distributions from a finite set of samples,''
\emph{Physical Review E}, vol. 52, no. 6, pp. 6841--6854, Dec. 1995,
doi: 10.1103/PhysRevE.52.6841.

\bibitem{Scutari2018}
M. Scutari,
``Dirichlet Bayesian network scores and the maximum relative entropy principle,''
\emph{Behaviormetrika}, vol. 45, no. 2, pp. 337--362, Oct. 2018,
doi: 10.1007/s41237-018-0048-x.

\bibitem{SchuermannGrassberger1996}
T. Sch\"urmann and P. Grassberger,
``Entropy estimation of symbol sequences,''
\emph{Chaos}, vol. 6, no. 3, pp. 414--427, Sep. 1996,
doi: 10.1063/1.166191.

\bibitem{WillemsShtarkovTjalkens1995}
F. M. J. Willems, Y. M. Shtarkov, and T. J. Tjalkens,
``The context-tree weighting method: Basic properties,''
\emph{IEEE Transactions on Information Theory}, vol. 41, no. 3, pp. 653--664, May 1995,
doi: 10.1109/18.382012.

\bibitem{GaoKontoyiannisBienenstock2008}
Y. Gao, I. Kontoyiannis, and E. Bienenstock,
``Estimating the entropy of binary time series: Methodology, some theory and a simulation study,''
\emph{Entropy}, vol. 10, no. 2, pp. 71--99, Jun. 2008,
doi: 10.3390/entropy-e10020071.

\bibitem{KontoyiannisEtAl2022BCT}
I. Kontoyiannis, L. Mertzanis, A. Panotopoulou, I. Papageorgiou, and M. Skoularidou,
``Bayesian context trees: Modelling and exact inference for discrete time series,''
\emph{Journal of the Royal Statistical Society: Series B (Statistical Methodology)}, vol. 84, no. 4, pp. 1287--1323, Sep. 2022,
doi: 10.1111/rssb.12511.

\bibitem{PapageorgiouKontoyiannis2023ITW}
I. Papageorgiou and I. Kontoyiannis,
``Truly Bayesian entropy estimation,''
in \emph{Proc. 2023 IEEE Information Theory Workshop (ITW)}, Saint-Malo, France, Apr. 2023, pp. 497--502,
doi: 10.1109/ITW55543.2023.10161645.

\bibitem{Lezaud1998}
P. Lezaud,
``Chernoff-type bound for finite Markov chains,''
\emph{The Annals of Applied Probability}, vol. 8, no. 3, pp. 849--867, Aug. 1998,
doi: 10.1214/aoap/1028903453.

\bibitem{HallHeyde1980}
P. Hall and C. C. Heyde,
\emph{Martingale Limit Theory and Its Application}.
New York, NY, USA: Academic Press, 1980.

\bibitem{MeynTweedie2009}
S. P. Meyn and R. L. Tweedie,
\emph{Markov Chains and Stochastic Stability}, 2nd ed.
Cambridge, U.K.: Cambridge Univ. Press, 2009,
doi: 10.1017/CBO9780511626630.

\end{thebibliography}
\end{document}